\documentclass{llncs}

\usepackage{natbib}
\usepackage{graphicx}

\usepackage{fancyhdr}
\usepackage{lastpage}
 
\begin{document}

\title{A short review and primer on using video for psychophysiological observations in human-computer interaction applications}
\author{Teppo Valtonen\inst{1}}
\institute{Quantified Employee unit, Finnish Institute of Occupational Health,\\
\email{teppo. valtonen @ttl. fi},\\
POBox 40, 00250, Helsinki, Finland}

\maketitle              % typeset the title of the contribution

\begin{abstract}

The application of psychophysiological measures in human-computer interaction is a growing field with significant potential for future smart personalised systems. Working in this emerging field requires comprehension of an array of physiological signals and analysis techniques.

An important aspect in measuring psychophysiological variables in real-world settings is the invasiveness of the measurement setup. Video is a signal which can be captured from a distance without interrupting the subject. Furthermore, the advancements in camera technologies enable detecting a growing variety of psychophysiological phenomena from a video signal with an increasing accuracy.

This paper aims to serve as a primer for the novice, enabling rapid familiarisation with the latest core concepts. We put special emphasis on everyday human-computer interface applications to distinguish from the more common clinical or sports uses of psychophysiology.

This paper is an extract from a comprehensive review of the entire field of ambulatory psychophysiology, including 12 similar chapters, plus application guidelines and systematic review. Thus any citation should be made using the following reference:\newline

{\parshape 1 2cm \dimexpr\linewidth-1cm\relax
B. Cowley, M. Filetti, K. Lukander, J. Torniainen, A. Henelius, L. Ahonen, O. Barral, I. Kosunen, T. Valtonen, M. Huotilainen, N. Ravaja, G. Jacucci. \textit{The Psychophysiology Primer: a guide to methods and a broad review with a focus on human-computer interaction}. Foundations and Trends in Human-Computer Interaction, vol. 9, no. 3-4, pp. 150--307, 2016.
\par}

\keywords{video, psychophysiology, human-computer interaction, primer, review, chapter}

\end{abstract}

\section{Introduction}
Advances in camera technologies make video an attractive possibility for measuring a variety of physiological phenomena, especially in environments such as the workplace, where contextual factors (e.g., ambient light) can be accounted for. Ever smaller and more accurate camera systems enable unobtrusive observations with sufficiently high precision for many interesting applications, and as long as there is a line of sight between the camera and the object of interest, a video signal may reveal, for example, a person's cognitive state without interrupting the work that is being done.

Here we consider digital video-based systems aimed at assessing some aspect of physiology or behaviour from a distance, in order to augment human--computer interaction. We exclude the ocular system from discussion, since it is dealt with in \cite{Cowley2016}.

\section{Background}
Typically, a video signal comprises measurements of the intensity of electromagnetic radiation in the spectra of visible (wavelengths of about 390 to 700 nm) and infrared (wavelengths from about 700 to 1000 nm) light, on a plane (for example, the image sensor in digital cameras). Changes in intensity arise mainly from a change in the original light source or the various points of reflection along the path of the ray of light from the source to the sensor. Accordingly, any movement within the measurement space (for instance, the eyebrows rising when the subject is surprised or expansion of the lungs when one is inhaling) or changes in the reflective properties of the reflection points (such as a change in skin colour due to increased blood flow) may be detected via the sensor. There is great variety in the video technologies and systems available today.

\paragraph{High-speed cameras}
A typical video camera captures 24 to 30 frames per second, depending on the encoding. While this is sufficient to make a video stream seem smooth for the human visual system, systems with higher frame rates have been developed too. One of the fastest methods, known as compressed ultra-fast photography (CUP), can capture non-repetitive time-evolving events at up to 10\textsuperscript{11} frames per second \citep{Gao2014}. For many physiological phenomena, a frame rate on the magnitude of 100 frames per second is adequate, and 200 frames a second may already allow, for example, the use of video-based photoplethysmography in clinical settings \citep{Sun2012a}.

\paragraph{Webcams}
The first system to feature a video camera that streamed an image in real time through a computer network came about in 1991 \citep{TrojanCoffeeCam}. The camera  was pointed at a coffee pot in the Cambridge University Computer Lab. Since then, video cameras have become a basic feature in laptop computers and the screens of desktop computers, and they have been used mainly for video calls.

\paragraph{Cameras in hand-held devices}
While cameras forming part of traditional computers are widespread, probably the most ubiquitous camera systems today are those embedded in hand-held devices, since almost all modern mobile phones and tablet computers have one or more cameras on their faces. The main camera typically points away from the user and is intended for photography. There is often another, however, intended for video calls and points in the same direction as the screen. In addition to conveying a video image to a caller, the front-facing camera can be used to detect, for example, whether or not there is a face in front of the screen. Once a face is detected, it may reveal various attributes of the user, such as emotional engagement as assessed from facial expressions \citep{Kang2008}. Also, however, as their name suggests, hand-held devices are often held in the user's hand. It has been demonstrated that the optical sensor of a mobile phone can detect, for example, the following elements from touch: breathing rate, heart rate, blood oxygen saturation, and even atrial fibrillation or blood loss \citep{Scully2012}.

\paragraph{3D camera systems}
Whereas a single-sensor camera system typically is limited to collecting emitted visual information on a two-dimensional plane, adding sensors and possibly projectors to the system may enable the observation of three-dimensional structures. A system with two appropriately placed cameras (i.e., a stereo camera system) functions in the same way as human binocular vision and can produce three-dimensional images. In another commonly used method, the system utilises projections of structured light patterns onto a three-dimensional surface, whereby the sensor's detection of distortions in the patterns received may reveal the precise three-dimensional co-ordinates of the surface \citep{Ma2009}. A third method, one that is quite new, uses a time-of-flight (ToF) camera system, in which distances from the source of a light pulse to a camera via each point in the visual field can be resolved from the time of flight of each light pulse on the basis of the known speed of light \citep{Gokturk2004}. Systems of this type have been used, for example, to monitor sleep \citep{Lee2015}.

\section{Methods}
With the above foundations laid, we now describe methods that have been used to extract information about human psychophysiology from a video signal. While there are diverse methods, we concentrate on three main categories here: light intensity analysis, 2D morphological analysis, and 3D morphological analysis. All of these areas are showing rapid development, and some solutions are still experimental. For practical methods, therefore, more research might be required.

\paragraph{Light intensity analysis} is the basis for video signal analysis and enables most higher-level interpretations. Even on its own, however, simply detecting changes in the intensity of the light in a fixed area in a video image may illuminate interesting psychophysiological variables. For example, while a plethysmograph reveals changes in volume in a body, typically due to changes in the amount of blood or air contained in that part of the body, photoplethysmography is an optical technique that can be used to detect variations in the intensity of light reflected from tissue that arise from changes in blood flow \citep{Allen2007}. For an optimal result, various light intensity parameters should be considered. These depend on the application. For example, pulses in line with heart rate seem the most apparent in the green colour channel of a colour camera feed \citep{Sun2012b}.

\paragraph{2D morphological analysis} is the analysis of interesting areas or shapes in a 2D image, and it is based on the detection of edges between areas that differ in light intensity. For HCI purposes, the most interesting part of the body is the human face. Face detection and recognition are established research topics, and there are free tools available for these (for example, \textit{OpenCV FaceRecognizer}\footnote{ See \url{http://docs.opencv.org/2.4/modules/contrib/doc/facerec/facerec\_tutorial.html}0.}). \citet{Samal1992} provide a good description of the process, from face detection all the way to the analysis of facial expressions and the classification of faces. More recently, \citet{Zhao2003} undertook an extensive review of face recognition. In a recent, thorough review, \citet{Martinez2016} concentrates on automatic recognition of facial expressions.

\paragraph{3D morphological analysis} is a broad category of analysis methods that rely on different optical sensor systems producing data on 3D structures within a sensor's field of view. For example, in addition to using intensity analysis, one can collect plethysmographic data from a distance by measuring the movement of a body in three-dimensional space with an optical sensor. Even consumer-grade 3D sensors used in gaming may be utilised to measure heart and respiration rate \citep{Bernacchia2014}.

\section{Applications}

Novel video technologies and methods of signal processing give rise to interesting applications for observing psychophysiological phenomena. Here we describe two of the most interesting video-based applications for HCI: photoplethysmography and the recognition of facial expressions.

Plethysmographic data can provide basic information on psychophysiology. More thorough description is given, for example, in \cite{Cowley2016}. Here we consider using video cameras for PPG, an optical technique that can be utilised to detect changes in blood flow \citep{Allen2007}. For example, Sun and colleagues extracted a PPG signal from a video and analysed pulse rate variability (a possible surrogate measure for HRV; see, for example, \citet{Gil2010}) from the palm of the subject's hand, using a monochrome CMOS camera running at 200 frames per second in 10-bit greyscale \citep{Sun2012a}. \citet{Tarassenko2014} measured both heart and respiration rate in a clinical set-up from a five-megapixel face video with eight bits per pixel, recorded at 12 frames per second. For broader applicability, even low-cost webcams have been demonstrated to function as photoplethysmographic sensors. In another of their studies, \citet{Sun2012b} compared a high-performance camera and a low-cost webcam in normal office lighting. They fixed a high-speed colour CMOS camera and a colour webcam in front of the face, along with a gold-standard pulse oximetry contact sensor on the index finger of the user's left hand. To ensure that they were measuring light reflected from the skin, the authors manually determined the region of interest (ROI) in each frame of the video signals. For HR detection, they used the green colour channel to analyse changes in the average intensity of the pixels within the ROI, since, especially in the webcam signal, pulsations were most apparent in this particular channel. They concluded that both imaging PPG systems can successfully measure important physiological variables (in their case, HR).

\begin{figure}[!t]
   \centering
   \includegraphics[scale=1.0]{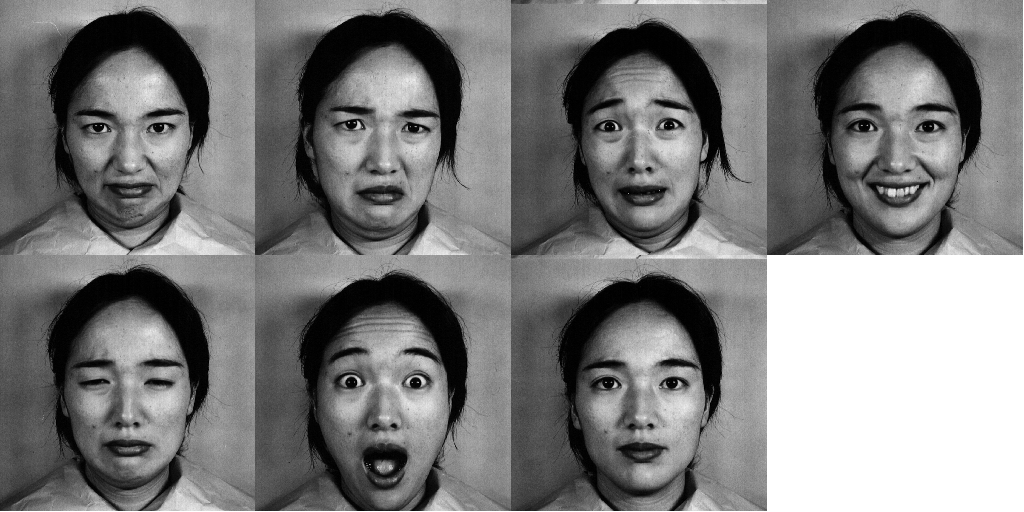}
   \caption{\textbf{Seven actor poses for emotional expressions, from the Japanese Female Facial Expression (JAFFE) database} \citep{Lyons1998}. Clockwise from top left: anger, disgust, fear, happiness, a neutral expression, surprise, and sadness. Although use of posed emotional expressions is a dated technique, such datasets provide a convenient picture of the concepts involved.}
   \label{fig.japfacs}
\end{figure}

Mental state is high-level information, and identifying and conveying that information is gaining burgeoning interest in HCI research. Knowledge of the user's mental state could augment not only user interfaces but also remote collaboration, telecommuting, and video conferencing. Facial expressions are an obvious signal as to the mental state, and the recognition of facial expressions is natural (and automatic) for humans; it is an important part of our communication. Both voluntary and involuntary facial actions convey, in particular, emotional information that is otherwise difficult to express -- and difficult to conceal in face-to-face interaction \citep{Ekman2003}. Facial expressions can be categorised as reflecting six canonical emotions in addition to a neutral expression: anger, disgust, fear, happiness, surprise, and sadness \citep{Dalgleish1999}. For an excellent review of automatic facial expression recognition, we recommend the work of \citet{Martinez2016}. Traditionally, the process has begun with databases of portraits of actors mimicking the emotions (for example, as shown in Figure~\ref{fig.japfacs}). A neutral expression is used as a reference for training the algorithm.

%...permission = creative commons

The core challenges with such a category-based approach are that the emotions each appear relatively rarely and that some expressions may differ in meaning on the basis of context. For example, someone might smile when embarrassed, not just when happy. Another method involves looking at the basic units of muscle activity in the human face, termed `action units', in keeping with the Facial Action Coding System (FACS; see \citep{Ekman1976}). With the FACS approach, interpretation of mental state can be done at a later stage in the analysis pipeline, with the aid of additional information on the context. A third approach is to represent the mental state on two or more dimensions of affect, such as continua for arousal (ranging from relaxed to aroused) and valence (from pleasant to unpleasant). However, neither action units nor values on the affective dimensions are always detected with the current methods, especially in real-world settings.

\section{Conclusion}

The recent progress in video and signal processing methods renders video an interesting alternative to many traditional means of obtaining psychophysiological measurements, in areas such as plethysmography. In addition, video may enable new HCI applications, such as the remote and automatic identification of the mental state. Many of these methods are still in the early phases of development and require more research before they can achieve greater feasibility; however, there is already inexpensive hardware available, and, for many areas of study, excellent-quality free, open-source software tools and libraries exist. It is clear that the use of video in HCI is only just beginning.

\bibliographystyle{plainnat}
\bibliography{ch9_video_bib}

\end{document}